\documentclass[12pt]{iopart}
\usepackage{iopams}
\usepackage{graphicx}
\usepackage{subfig}
\usepackage{siunitx}
\usepackage{braket}
\usepackage{cite}
\usepackage[]{color}
\usepackage{upgreek}
\usepackage{float}
\usepackage{hhline}

\begin{document}


\title[A trapped ultracold atom force sensor with a $\mu$m-scale spatial resolution]{A trapped ultracold atom force sensor with a $\upmu$m-scale spatial resolution}

\author{X. Alauze$^1$, A. Bonnin$^1$, C. Solaro$^1$\footnote{Present address: Department of Physics and Astronomy, Aarhus University, DK-8000 Aarhus C, Denmark} and F. Pereira Dos Santos$^1$}

\address{$^1$ LNE-SYRTE, Observatoire de Paris, Universit\'e PSL, CNRS, Sorbonne Universit\'e, 61 avenue de l'Observatoire 75014 Paris, France}

\ead{franck.pereira@obspm.fr}

\begin{abstract}

We report on the use of an ultracold ensemble of $^{87}$Rb atoms trapped in a vertical lattice as a source for a quantum force sensor based on a Ramsey-Raman type interferometer.
We reach spatial resolution in the low micrometer range in the vertical direction thanks to evaporative cooling down to ultracold temperatures in a crossed optical dipole trap.
In this configuration, the coherence time of the atomic ensemble is degraded by inhomogeneous dephasing arising from atomic interactions.
By weakening the confinement in the transverse direction only, we dilute the cloud and drastically reduce the strength of these interactions, without affecting the vertical resolution.
This allows to maintain an excellent relative sensitivity on the Bloch frequency, which is related to the local gravitational force, of $5\times10^{-6}$ at 1\,s which integrates down to $8\times10^{-8}$ after one hour averaging time.

\end{abstract}
\pacs{37.10.Jk, 37.25.+k, 32.80.Qk, 05.60.Gg}

\noindent{\it Keywords\/}: atom interferometry, lattice, quantum sensor, atomic coherence, atomic interactions


\maketitle


\section{Introduction}

Atom interferometry has led to extremely sensitive and accurate inertial sensors such as gravimeters \cite{Gillot2014,Freier2016,Hu2013}, gradiometers\cite{Sorrentino2014} or gyrometers\cite{Dutta2016}.
These sensors are of great interest to perform tests of fundamental physics such as measuring fundamental constants \cite{Fixler2007,Rosi2014,Thomas2017,Bouchendira2011}, testing the equivalence principle \cite{Zhou2015,Bonnin2015,Tarallo2014,Schlippert2014}, detecting gravitational waves \cite{Geiger2015,Dimopoulos2008} or probing short range forces \cite{Dimopoulos2003,Wolf2007,Ivanov2012}.
Trapped atom interferometers in particular, allowing for longer interrogation times and thus for a better measurement sensitivity without increasing the interrogation spatial area, are paving the way for much more compact sensors. Moreover they provide better spatial resolution when compared to free falling atoms which is a key feature for short range forces measurements.
In this context, using as a test mass atomic clouds featuring at the same time the smallest size, for better spatial resolution, and the largest number of atoms, for optimized signal to noise ratio, leads to a regime of high densities, where atomic interactions become an important issue.
Such interactions induce mechanisms that can be either detrimental (inhomogeneous dephasing, collisional shifts \cite{Harber2002,Kuhr2005}) or favorable (spin-self rephasing \cite{Deutsch2010,Buning2011}, spin squeezing \cite{Gross2010}, entanglement \cite{Klempt2010,Lange2017,Berrada2013}) to the measurement and can lead to complex spin dynamics during the interferometric sequence \cite{Solaro2016}.

In the experiment reported here, where tens of thousands ultracold $^{87}$Rb atoms are trapped in only a few wells of a shallow 1-D vertical optical lattice, we explore high atomic densities ranging from $10^{10}$ to $10^{12}$ atoms/cm$^3$.
An interferometer in a Ramsey-Raman configuration \cite{Beaufils2011,Pelle2013} allows to probe the energy difference between adjacent lattice sites consequently providing a local measurement of the vertical force. 
Far from any source mass, the sensor measures the Earth gravitational acceleration.
Close to the surface of the retroreflecting mirror creating the lattice potential, this sensor will allow for a very sensitive measurement of short range forces, and more specifically of the Casimir-Polder force with an expected relative uncertainty better than the percent.

In a previous study \cite{Hilico2015} we have demonstrated a relative sensitivity of $3.9\times10^{-6}$ at 1\,s in the measurement of the Bloch frequency, which got later improved down to $1.8\times10^{-6}$ at 1\,s \cite{SolaroThesis}, comparable to the best ever reported sensitivity for a trapped cold atom force sensor of $1.5\times10^{-6}$ at 1\,s \cite{Tarallo2014}.
Yet, this interferometer had been performed with a large and diluted cloud loaded into the lattice from an optical molasses.
With an atomic cloud of about 2\,mm size, spreading over thousands of wells, this interferometer did not offer the spatial resolution required for short range measurements.

In this paper, we report on the use of all-optical evaporative cooling to reach a thousand time better spatial resolution, while preserving high sensitivity. The size of the atomic sample is reduced to the order of a few microns and the temperature is also decreased from $\sim$\,2\,$\upmu$K to $\sim$\,100\,nK. With atomic densities in the $10^{10}$ to $10^{12}$ atoms/cm$^3$ range, inhomogeneous dephasing induced by interactions constitute a limitation for the coherence of the atomic ensemble.
In the following, we report on the precise evaluation of its impact onto the measurement sensitivity, and demonstrate that it can be mitigated by optimizing the experimental parameters, without compromizing the spatial resolution.

After a brief description of the experimental setup and measurement principle, we present, for different trap parameters, contrast measurements as a function of the number of atoms, and show that coherence loss can be mitigated by reducing the confinement in the horizontal directions.
We finally show how the sensitivity of the measurement can be optimized. For that purpose, we carry out a detailed analysis of all relevant sources of noise and finally determine the optimal  interferometer parameters.

\section{The Experiment}

\subsection{Principle}\label{sec:principle}

\begin{figure}[!ht]
  \centering
  \includegraphics[width=0.6\textwidth]{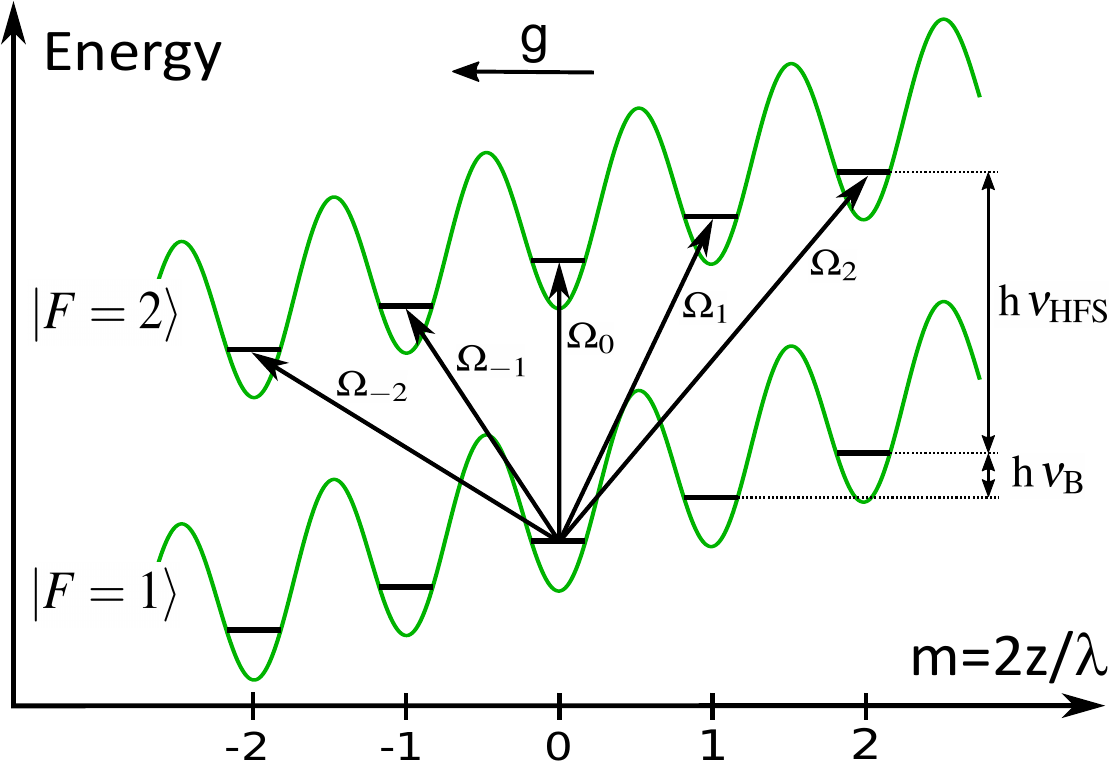}
  \caption{Ladders of Wannier-Stark states for a two level atom. We consider here the example of the two hyperfine ground states of an alkali atom, separated in frequency by $\nu_{\mathrm{HFS}}$. Adjacent wells are separated by the Bloch frequency $\nu_\mathrm{B}$. Raman laser pulses allow to couple neighbouring states, with Rabi frequencies $\Omega_{\Delta_m}$, which depend on the absolute distance between the wells.}
  \label{WSladder}
\end{figure}

The system has been described in detail in previous work \cite{Hilico2015,Pelle2013,Zhou2013,Tackmann2011,Beaufils2011}.
$^{87}$Rb ultra-cold atoms are trapped in a shallow 1D vertical optical lattice.
This system features localized pseudo-eigenstates, which compose the so-called ladder of Wannier-Stark states \cite{Wannier1960,Gluck2002} $\ket{WS_m}$, where the quantum number $m$ is an index labeling the lattice sites.
The energy separation between two consecutive Wannier-Stark states is simply given by the difference in gravitational potential between two consecutive wells :
\begin{equation}
h\,\nu_\mathrm{B} = m_{Rb}\,g\,\frac{\lambda_l}{2}
\end{equation}
where $\lambda_l$ = 532\,nm is the lattice laser wavelength and $\nu_\mathrm{B} \approx$ 568.5\,Hz is the Bloch frequency \cite{Zener1934}, $m_{Rb}$ is the atomic mass, g the acceleration of gravity.
Taking into account the two internal states $\ket{f} = \ket{5\ ^2S_{1/2},F=1,m_F=0}$ and $\ket{e} = \ket{5\ ^2S_{1/2},F=2,m_F=0}$, one obtains two Wannier-Stark ladders of states separated by an energy $h\nu_{\mathrm{HFS}}$ where $\nu_{\mathrm{HFS}} \approx \SI{6.835}{\giga\hertz}$ is the hyperfine structure frequency. The WS ladders for the two sets of eigenstates $\ket{f,WS_m}$ and $\ket{e,WS_m}$ are shown in \fref{WSladder}.

For shallow lattice depths ($U_0 < $ 10\,E$_{rec}$, where E$_{rec}$ is the recoil energy of a lattice photon), the atomic wave function spans across several wells, allowing for a laser induced coherent tunneling between different lattice sites. Resonant two-photon Raman transitions, using two counter-propagating beams, can be performed to couple the two states $\ket{g,WS_m}$ and $\ket{e,WS_{m+\Delta m}}$ in the same well ($\Delta m=0$) or in different wells ($\Delta m\neq 0$).
When the frequency difference between the two Raman lasers yields the resonance condition $\nu_{\mathrm{Raman}} = \nu_{\mathrm{HFS}} + \Delta m \times \nu_\mathrm{B}$, we selectively address transitions between states separated by $\Delta m$ wells of the lattice (see \fref{WSladder}). The coupling is optimized by tuning the lattice depth to adjust the WS state delocalization for a chosen separation $\Delta m$ \cite{Beaufils2011,Tackmann2011}.

\subsection{Atomic source}\label{sec:preparation}

The lattice is loaded with ultracold atoms produced according to the following preparation sequence.
First, $1.5\times10^9$ atoms are trapped in a 3 dimensional magneto-optical trap (3D-MOT), loaded within 600\,ms from a 2D-MOT.
The cloud is then compressed for 100 ms in a dark MOT, by lowering the intensity of the repumper laser and increasing both the magnetic field gradient and the detuning. After turning off the MOT, about $10^7$ atoms are transferred into an optical dipole trap, predominantly in the $\ket{F=1}$ state.
Two beams of a high power Yb fiber laser at 1070 nm are crossed in the horizontal plane, with an angle of 43$^{\circ}$. They are switched on 100 ms before the MOT is turned off and focused onto the atoms with $50 \upmu$m and $70 \upmu$m radii at $1/e^2$ and maximum powers of respectively 10 and 20 W. The power is then exponentially ramped down to 0.12 and 0.23 W within 1.25\,s, which yields fast evaporative cooling.
A sample of $\sim10^5$ atoms at a temperature of $\sim$300 nK is obtained. The final phase space density is 0.5, close to degeneracy.

At the beginning of the evaporation, a vertical bias field of about 70 mG is applied and the atoms are optically pumped into the $\ket{F=1,m_F=0}$ state with 70 \% efficiency with a 1.2 ms long pulse of horizontally linearly polarized light tuned on the $\ket{F=1} \rightarrow \ket{F'=0}$. We attribute this imperfect pumping to the absorption of the pumping light by the optically thick sample of atoms. In order to purify the polarization of the sample, a sequence of microwave and pusher pulses is then used at the end of the evaporation. A first microwave pulse transfers the atoms from $\ket{F=1, m_F=0}$ into $\ket{F=2, m_F=0}$ and a subsequent 12 ms long pulse of optical pumping heats up the atoms remaining in $\ket{F=1}$, which escape from the trap. A second microwave pulse transfers the atoms from $\ket{F=2, m_F=0}$ back into $\ket{F=1, m_F=0}$. The small fraction of atoms remaining in $\ket{F=2}$ (about 3\%) is finally pushed and more than 99.5\% of the atoms are in the state $\ket{F=1, m_F=0}$ which is insensitive, to first order, to magnetic fields.

The atoms are then adiabatically transferred, within 100\,ms, into the vertical optical lattice created with a retro-reflected laser beam at 532\,nm. With a 500\,$\upmu$m waist and a power of 6\,W, the maximal lattice depth is 7\,E$_{rec}$.
This standing wave being blue-detuned, the atoms are not trapped in the transverse direction. We therefore superimpose a red-detuned progressive wave created by a laser beam at 1064\,nm, with a waist ranging from 150 to 300 $\upmu$m and a power of up to 2\,W, to constrain the atoms at the maximum intensity of the lattice.
Finally, we end up in this trap with a maximum number of atoms of a few $10^4$ at transverse temperatures in the range 50-200 nK, depending on the power and waist of the radial confinement laser, which corresponds to atomic densities in the $10^{11}-10^{12}$  at/cm$^3$ range (about three orders of magnitude higher than in previous configurations \cite{Hilico2015}, where the cloud was loaded from an optical molasses).

\subsection{Interferometer schemes and measurement}\label{sec:interfero}

We describe in this section our interferometer geometry. Atoms, initially in the state $\ket{F=1, m_F=0, WS_m}$, are coupled to the state $\ket{F=2, m_F=0, WS_{m+\Delta m}}$ via a two-photon Raman transition (see \sref{sec:principle}).
The counterpropagating Raman lasers are phase locked onto an ultra low noise reference oscillator. They have $\sigma^+$ - $\sigma^+$ polarizations, identical waists of about 1 mm and powers of a few mW. They are detuned from the D2 transition by 300 GHz to avoid loss of coherence due to spontaneous emission and to reduce differential light shift (DLS) inhomogeneities. Typical Rabi frequencies are of the order of $\Omega_{eff} \sim 2\pi \times 25$ Hz, corresponding to a $\pi/2$ pulse duration $\tau_{\pi/2} \sim 10$ ms.

\begin{figure}[!ht]
    \centering
    \subfloat[Symmetrized Ramsey-MicroWave Interferometer (SRMWI).]{{\includegraphics[width=0.255\textwidth]{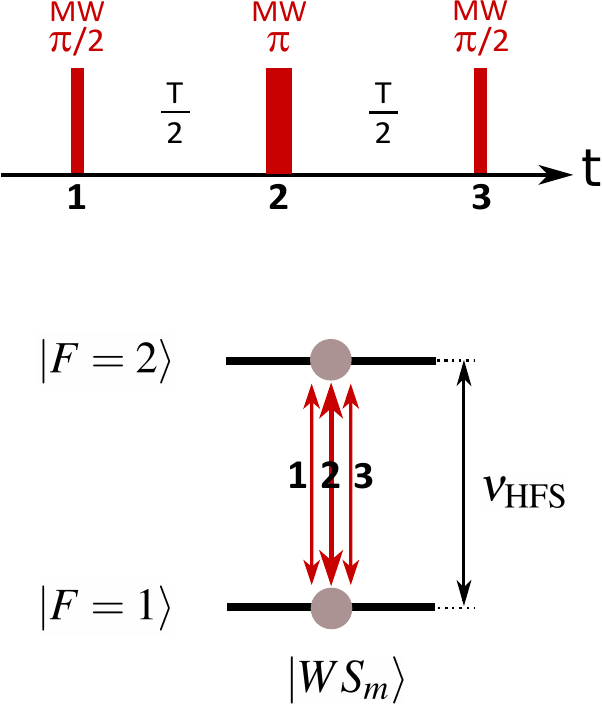} }\label{RMWS}}
    \;\;\;\;\;\;\;
    \subfloat[Ramsey-Raman Interferometer (RRI). $\Delta\phi\propto \nu_{\mathrm{HFS}}+\nu_{\mathrm{B}}$]{{\includegraphics[width=0.30\textwidth]{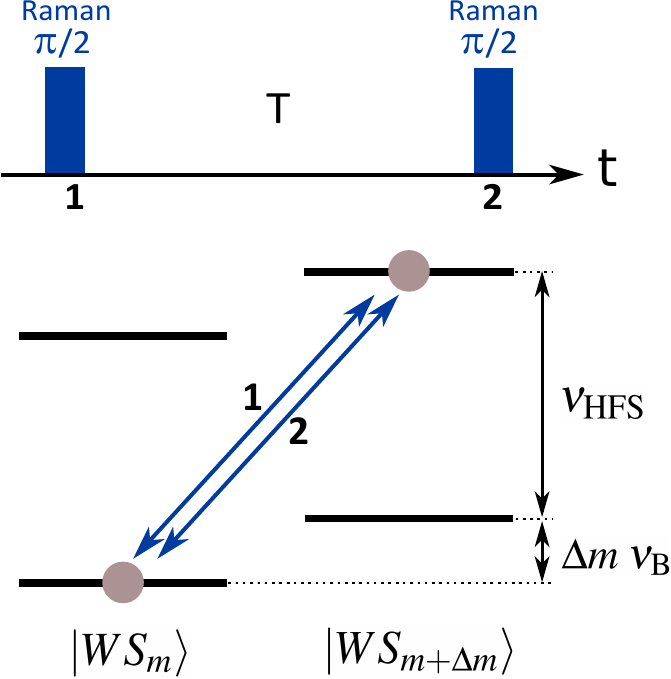} }\label{RR}}
    \;\;\;\;\;\;
    \subfloat[Symmetrized Ramsey-Raman Interferometer (SRRI). $\Delta\phi\propto\nu_{\mathrm{B}}$]{{\includegraphics[width=0.30\textwidth]{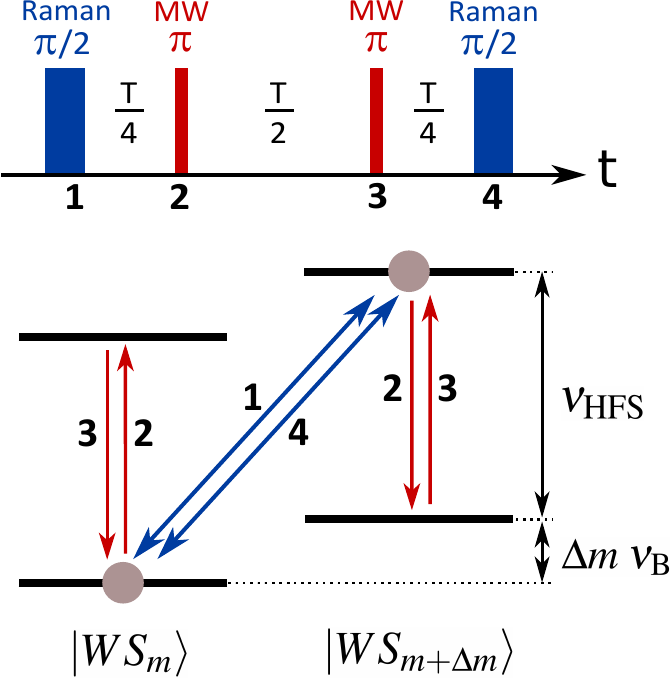} }\label{RRSMW}}
    \caption{Raman and microwave (MW) sequence of pulses along with the relevant energy levels for three different interferometer schemes: (a) Symmetrized Ramsey-Microwave, (b) Ramsey-Raman and (c) Symmetrized Ramsey-Raman.}
    \label{Interfero}
\end{figure}

A sequence of two $\pi/2$ pulses, acting as a beamsplitter and a recombination pulse, and separated by a free evolution time $T$, allows to create a Ramsey-Raman interferometer \cite{Pelle2013} with separated spatial states (see \fref{RR}). The phase shift of the interferometer is proportional to the difference in energy between the two states separated by $\Delta m$ wells:
\begin{equation}
\Delta\phi_{RRI}(\Delta m)	= 2\pi\times(\nu_{\mathrm{Raman}}-\nu_{\mathrm{HFS}}-\Delta m\ \nu_\mathrm{B}) \times T
\label{PhaseShiftRR}
\end{equation}
where $\nu_{\mathrm{Raman}}$ is the detuning between the two Raman lasers.

This interferometer is at the same time a clock and an inertial sensor: it is sensitive to inertial effects through the dependence of this phase on the Bloch frequency and to clock shifts through its dependence on the hyperfine frequency.
When used as a force sensor, it will thus be biased by frequency shifts of the clock transition, such as due to the DLS induced by the trapping lasers \cite{Hilico2015,Kuhr2005} and the cold collision shift induced by atomic interactions \cite{Sortais2000}. More, dephasing due to frequency shift inhomogneities lead to contrast loss at long interrogation times and degrades the sensitivity of the force measurement. To suppress, at least partially, the impact of these frequency shifts, we realize a symmetric version of the interferometer, as displayed in \fref{RRSMW}, where two additional microwave $\pi$ pulses are inserted in between the two Raman pulses in order to exchange the two internal states without changing the external states, so that the spatially separated wavepacket now spend the same time in each of the two hyperfine states.
This symmetrization procedure reduces dephasing, which improves the contrast \cite{SolaroThesis}, and makes the interferometer phase independent of the hyperfine structure splitting. This phase is now given by:
\begin{equation}
\Delta\phi_{SRRI}(\Delta m)	= 2\pi\times(\nu_{\mathrm{Raman}}-\nu_{\mathrm{MW}}-\Delta m\ \nu_\mathrm{B}) \times T
\label{PhaseShiftRRSMW}
\end{equation}
where $\nu_{\mathrm{MW}}$ is the frequency of the microwave source.

Clock type Ramsey interferometers can also be performed, with no separation, using only two MW pulses.
The interferometer phase shift is then:
\begin{equation}
\Delta\phi_{RMWI}	= 2\pi\times(\nu_{\mathrm{MW}}-\nu_{\mathrm{HFS}}) \times T
\label{PhaseShiftRMW}
\end{equation}
We can also symmetrize this interferometer with a third MW pulse as shown in \fref{RMWS}. This configuration is used to separate noise contributions from the Raman laser contribution. It is insensitive to the clock transition and the interferometer fringes are recorded by scanning the phase of the last MW pulse.

\bigbreak
We exploit the state labelling of the two-photon Raman transitions \cite{Borde1989,Pelle2013} to read out the interferometer phase.
Fluorescence measurements of the populations $N_1$ and $N_2$ in the two internal states $\ket{F=1}$ and $\ket{F=2}$ allow to extract the interferometer phase via the measurement of the transition probability given by
\begin{equation}
P	=  \frac{N_2}{N_1+N_2} = \frac{1}{2}[1-C\,\mathrm{cos}(\Delta\phi)]
\label{TransitionProba}
\end{equation}
where $C$ is the contrast of the interferometer.
This normalized detection scheme makes the transition probability insensitive to atom number fluctuations.

The measurement of the Bloch frequency $\nu_\mathrm{B}$ is realized with a symmetrized Ramsey-Raman interferometer (see \fref{RRSMW}). A digital lock is performed \cite{Pelle2013} to operate at mid-fringe of the central fringe of the interferometer, where the sensitivity is maximal.
We interleave frequency measurements with separations of $\pm \Delta m$ wells (which have identical Raman coupling, but different Raman resonance conditions). Measurements of the Bloch frequency are finally derived from the difference of frequencies between the two interleaved configurations. This measurement procedure allows suppressing common mode systematic frequency shifts and improving the long term stability.

\section{Optimization of the sensitivity}\label{sec:sensitivity}

From equation \eref{PhaseShiftRRSMW} and \eref{TransitionProba} and for $\Delta\phi \to \frac{\pi}{2}$ (where the sensitivity is maximal), the relative sensitivity on the Bloch frequency at 1\,s measurement time is given by:
\begin{equation}
\frac{\sigma_{\nu_B}}{\nu_B}	= \frac{1}{\Delta m\,\nu_B}\ \frac{\sigma_P}{\pi\,C\,T}\sqrt{T+T_p}
\label{sigma_nu}
\end{equation}
for a separation of  $\Delta m$ wells, with $\sigma_P$ the standard deviation of shot to shot fluctuations of the transition probability and $T_p = \SI{2.3}{\second}$ the dead time, corresponding to the preparation and detection phases.

The larger the separation $\Delta m$, the better the sensitivity, but increasing the separation requires to lower the lattice depth to optimize the Raman coupling and reduce the coupling inhomogeneities. For separations $\Delta m \geq 7$, lattice depth below 1.3\,E$_{rec}$ are required, for which the number of atoms drops drastically, due to the exponential scaling of losses via Landau Zener tunelling \cite{Zener1932}, resulting in a detrimental increase in detection noise. We therefore use in the following a maximum separation of $\Delta m = 6$. 
To optimize the coupling, the lattice laser power is then set to 1.2 W which corresponds to a depth of 1.9\,E$_{rec}$.

The three other parameters impacting the sensitivity in equation \eref{sigma_nu} are $C$, $\sigma_P$ and $T$.
For small enough interferometer phase noise, $\sigma_P$ will be dominated by detection noise, which depends only on the number of detected atoms.
As for the contrast, it decreases with interrogation time due to the finite coherence time of the interferometer.
Optimizing the sensitivity thus results from a compromize between good contrast - but poor scale factor - at small $T$ and large scale factor - but poor contrast - for large $T$.
In our case, as we show later, the contrast decay rate depends on the number of atoms, which makes the optimum search multi-parameter.

\subsection{Contrast decay}\label{sec:ContrastDecay}

\begin{figure}[!ht]
    \centering
    \subfloat[Typical fringes for the symmetrized Ramsey-Raman interferometer. The decay rate increases with the density proportional to the number of atoms. Solid black: 7000 atoms  --  Dashed red: 21000 atoms.]{{\includegraphics[width=0.42\textwidth]{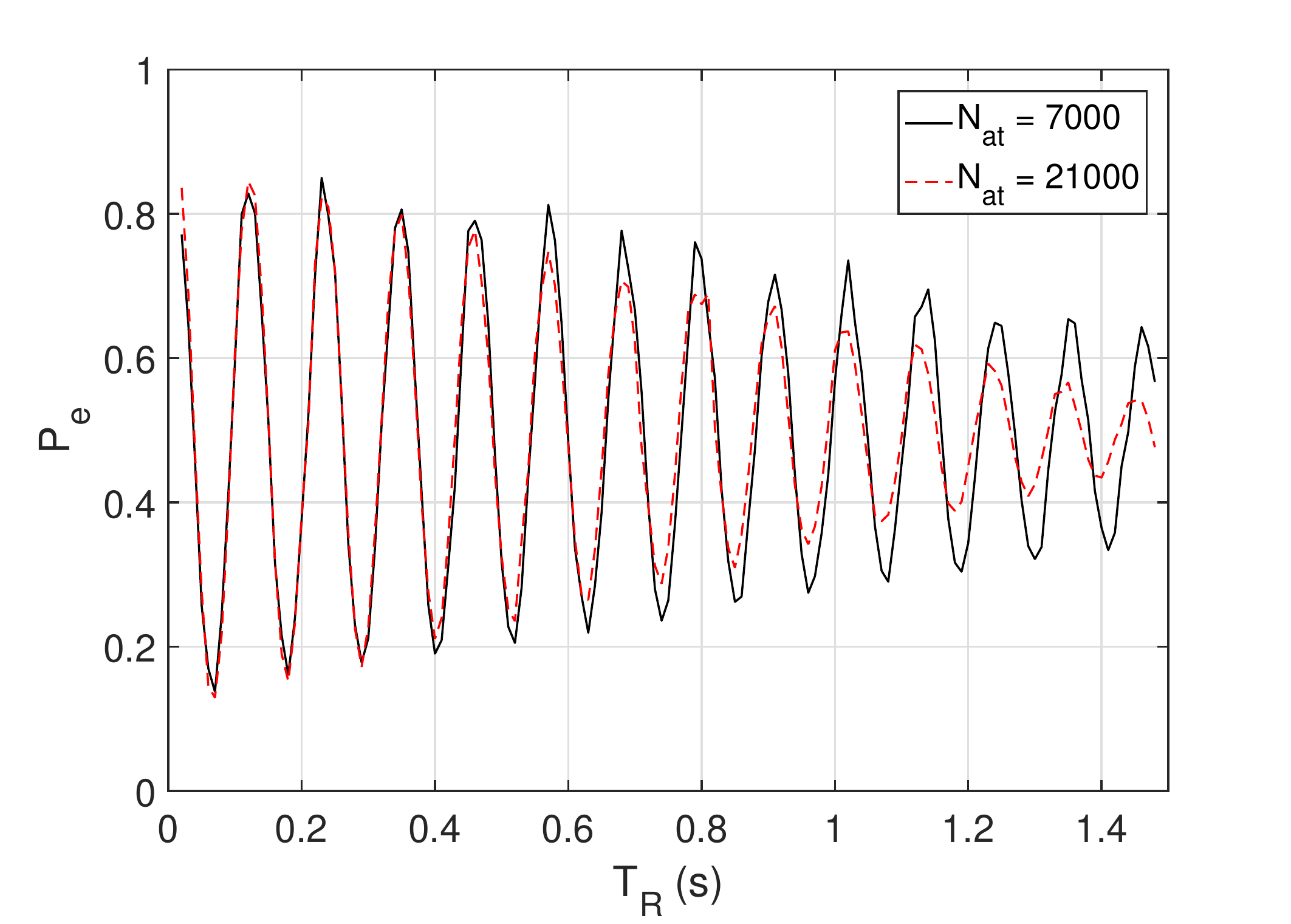} }\label{decayA}}
    \;\;\;\;\;
    \subfloat[Contrast decay rate $\gamma$  as a function of the number of atoms for three different configurations of waists and power. The contrast decays as: $C(T,N_{at}) = C_0\; e^{-\gamma(N_{at}) T}$. Solid lines are linear fits: $\gamma(N_{at}) = \gamma_0 + \alpha\, N_{at}$.]{{\includegraphics[width=0.51\textwidth]{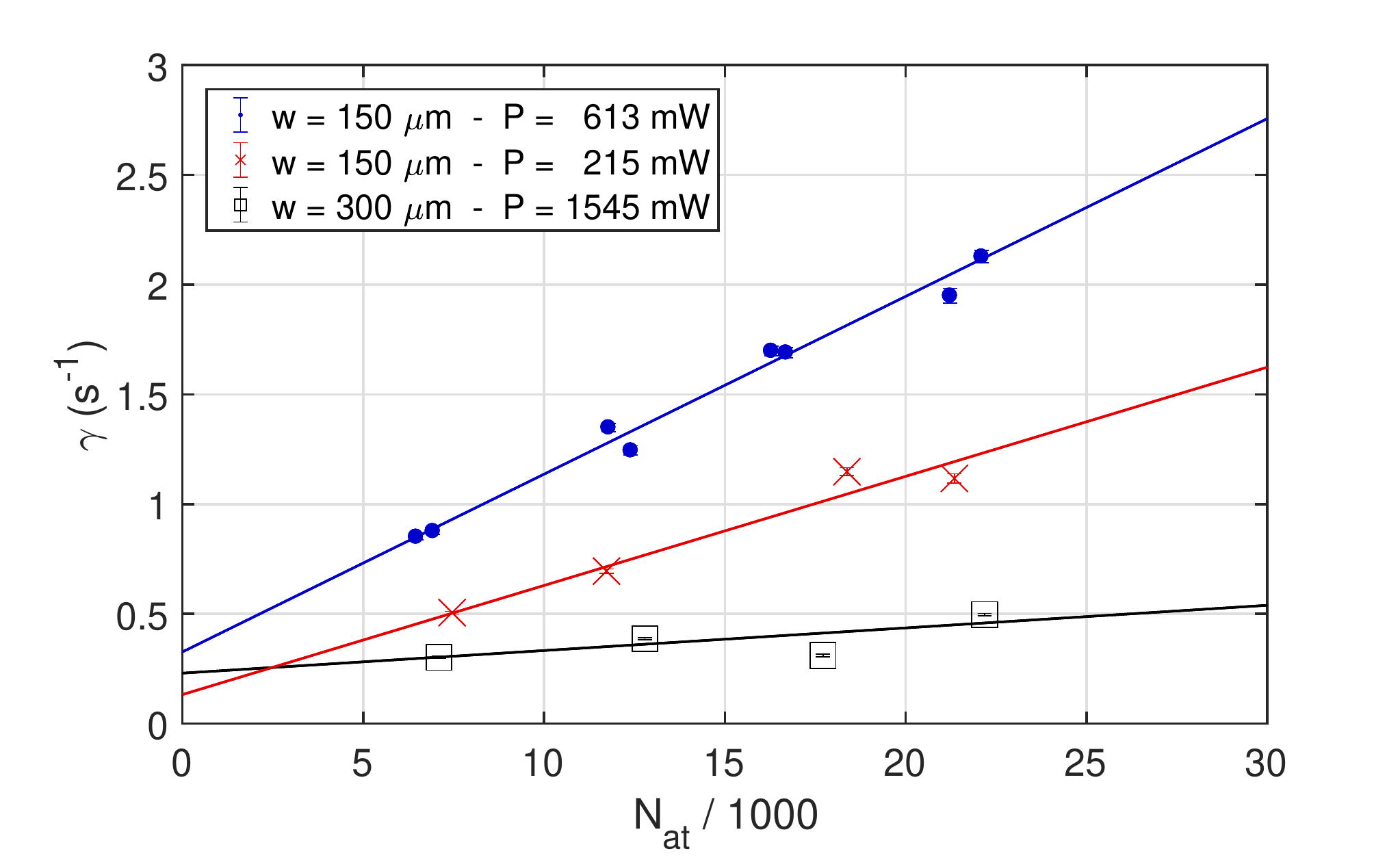} }\label{decayB}}
    \caption{Contrast decay of the Symmetrized Ramsey-Raman Interferometer (SRRI)}
    \label{decay}%
\end{figure}

To determine the contrast loss rate, we record the interferometer fringes by increasing T while scanning the phase difference between the two laser pulses, with microwave and Raman frequencies fixed on resonance. Such a fringe decay signal is displayed on \fref{decayA}.
The contrast as a function of time is extracted from the Ramsey fringes using the standard deviation of a sinusoidal function over one period and fitted with the function $C(T)=C_0\,e^{-\gamma T}$, where $\gamma$ is the contrast decay rate.

\Fref{decayB} displays measured decay rates as a function of the number of atoms $N_{at}$ for different parameters of the transverse confinement laser. $N_{at}$ is varied by changing the duration of the second MW pulse during the preparation (see \sref{sec:preparation}). We verified that, for each set of transverse confinement parameters,this selection method leaves the temperature and cloud volume unchanged, ensuring that the density scales linearly with the number of atoms.
We observe an increase of the contrast decay rates with $N_{at}$, which indicates that the dephasing due to atomic interactions is not completely suppressed by the symmetrization.
The effect of the symmetrization for this Ramsey-Raman interferometer with seperated arms is in noticeable contrast with the behaviour observed in Ramsey-MW interferometers (with no spatial separation), for which, in such regimes of densities, exchange collisions have been shown to lead to spin synchronization and spin self rephasing (SSR) \cite{Deutsch2010} and where special attention must be paid, when symmetrizing the interrogation pulses sequence, to the joint effects of SSR and the symmetrization pulses \cite{Solaro2016}. Here, contrary to \cite{Deutsch2010,Solaro2016}, the two wavepackets being spatially separated ($\Delta m \neq 0$), the exchange collision rate and hence the SSR are reduced. We thus observe neither the extended coherence times of \cite{Deutsch2010,Solaro2016} nor any non-monotonic behavior of the contrast due to the symmetrization pulses.

Fitting the data from \fref{decayB} with the function $\gamma(N_{at})=\gamma_0+\alpha\,N_{at}$ allows to distinguish interaction effects from other decoherence sources, such as related to the trapping lasers.
The extrapolated dephasing rate $\gamma_0$ for $N_{at}$ = 0 can be attributed to imperfect suppression of DLS (due for instance to laser intensity and pointing fluctuations, or to temperature changes caused by heating of the atomic cloud or residual evaporation) and to inhomogeneities in the parasitic dipolar forces if the atoms are not perfectly placed at the waist of the transverse confinement laser \cite{Hilico2015}. 
Given that interaction effects scale linearly with density, $\alpha$ is expected to be inversely proportional to the volume of the atomic cloud.
The decoherence induced by the interactions, the DLS and the parasitic dipolar force all depend on the power $P$ and the waist $w$ of the transverse confinement laser. Decreasing the power of the transverse confinement laser allows to reduce these three deleterious effects together. Nevertheless lowering the trap depth below $\sim 1 \upmu$K reduces drastically the number of trapped atoms. Consequently, we chose, for a given trap depth, to increase the waist and the power of the beam in order to dilute the cloud further and decrease significantly the effect of the interactions.

The blue points and the red crosses in \fref{decayB} correspond to measurements done with the same waist of 150\,$\upmu$m and different powers of respectively 613 and 215\,mW. The trap is deep enough in both cases for all the atoms to be transferred from the dipole trap after the evaporation. As expected, both the offset $\gamma_0$ and the slope $\alpha$ are both reduced.
The black squares correspond to a waist of 300\,$\upmu$m, for which the density (for a given number of atoms) is much lowered, thus corresponding to a reduced slope $\alpha$.

\begin{figure}[!ht]
  \centering
  \includegraphics[width=0.6\textwidth]{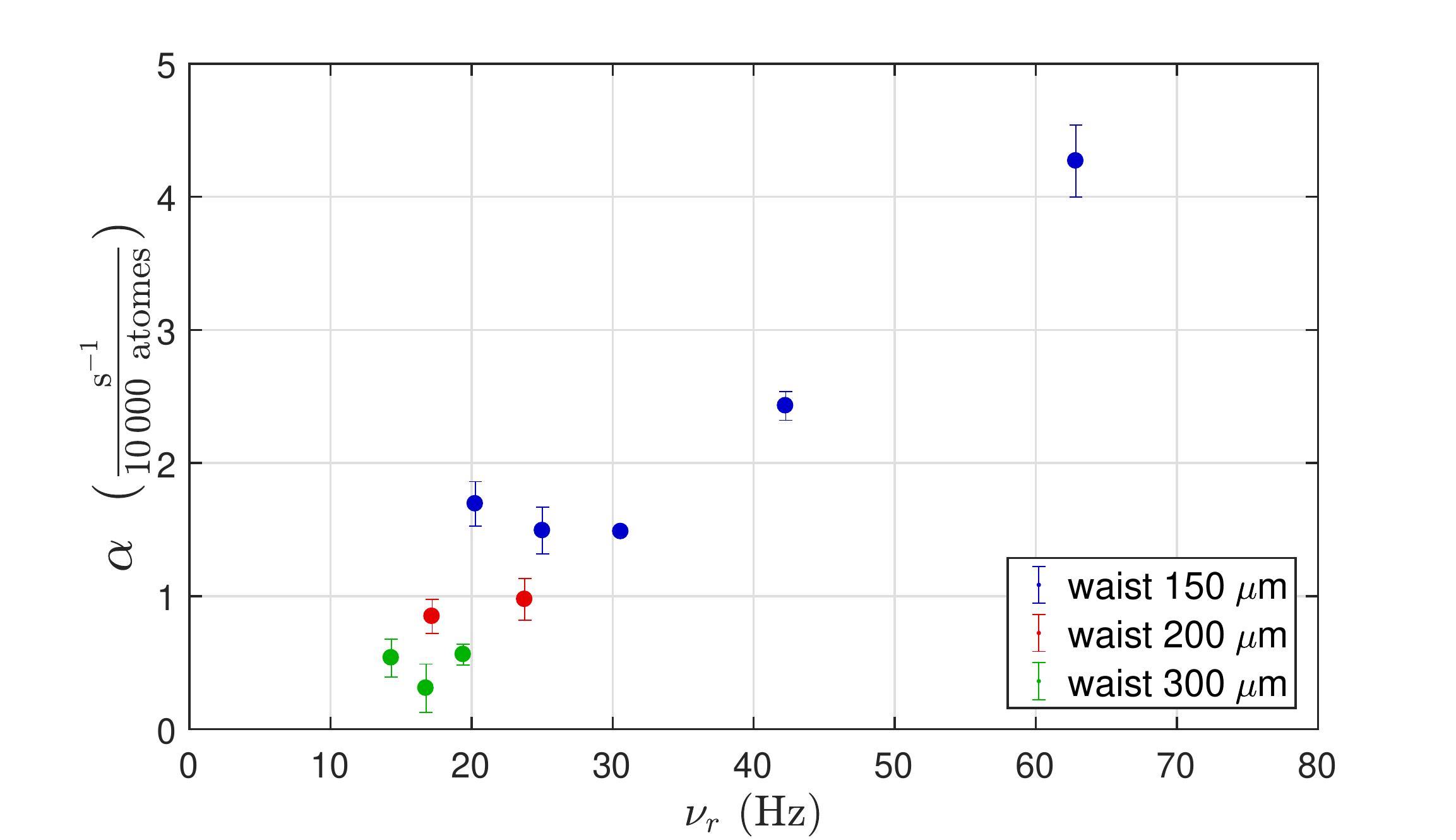}
  \caption{Parameters $\alpha$ extracted from the fit of the contrast decay rate $\gamma(N_{at}) = \gamma_0 + \alpha\,N_{at}$, as a function of the trap frequency $\nu_r$.}
  \label{alpha} 
\end{figure}

We repeated such measurements for different intensities, in the 6-60\,W/mm$^2$ range, and waists in the 150-300\,$\upmu$m range. \Fref{alpha} displays the parameter $\alpha$ extracted from these fits, as a function of the transverse trap frequency $\nu_r$. Assuming that the atoms are loaded adiabatically in the transverse potential and well within the harmonic regime (the trap depth ranges from 5 to 20 times the atomic cloud temperature), the volume of the trap, and thus $\alpha$, is expected to scale linearly with $\nu_r$. For low trap depths, the harmonic approximation is not valid anymore which explains why $\alpha$ does not increase linearly at small $\nu_r$.

In light of this analysis, we choose for the rest of the study to increase the waist of the transverse confinement laser from 150 $\upmu$m to 300\,$\upmu$m. The transverse size of the cloud is then increased from 36\,$\upmu$m to 74\,$\upmu$m and the density is divided by a factor 4. The contrast decay rate, at large atom number, is reduced by approximately a factor 4. With 15000 atoms and a power of 1.36\,W, the coherence time of the atomic sample is finally 1/$\gamma$ = 2.5\,s.

\subsection{Fluctuations of the transition probability}\label{sec:sigmaP}
To optimize the relative sensitivity \eref{sigma_nu} and choose the optimal parameters for the symmetrized Ramsey-Raman interferometer, we need to characterize the shot to shot fluctuations of the transition probability $\sigma_P$. We quantify below the contributions to $\sigma_P$ arrising from detection noise $\sigma_{det}$ (which depends only on the number of atoms), and from the interferometer phase noise $\sigma_{\phi}$ (which depends on the interferometer duration T). Taking those two contributions into account, $\sigma_P$ is expressed as:
\begin{equation}
\sigma_P^2(N_{at},T) = \sigma_{det}^2(N_{at})+\frac{C^2(N_{at},T)}{4}\ \sigma_{\phi}^2(T)
\label{sigmaP_mes}
\end{equation}

\subsubsection*{Detection noise}
\leavevmode\par
The detection noise $\sigma_{det}$ is given by:

\begin{equation}
\sigma_{det}^2 = \frac{a^2}{N_{at}^2}+\frac{1}{4N_{at}}+b^2
\label{sigma_det}
\end{equation}
where the first contribution is related to electronics noise (such as related to digitization noise, background light or voltage noise of the transimpedance circuit), the second to quantum projection noise and the third to technical noise (such as related to normalization noise or detection laser intensity and frequency noise). A detailed characterization of our detection scheme, with the atoms prepared in equal superposition of the $\ket{F=1}$ and $\ket{F=2}$ states using a single $\frac{\pi}{2}$ MW pulse, gives $a=58$ and $b=10^{-3}$.

\subsubsection*{Phase noise}
\leavevmode\par

\begin{figure}[!ht]
  \centering
  \includegraphics[width=0.6\textwidth]{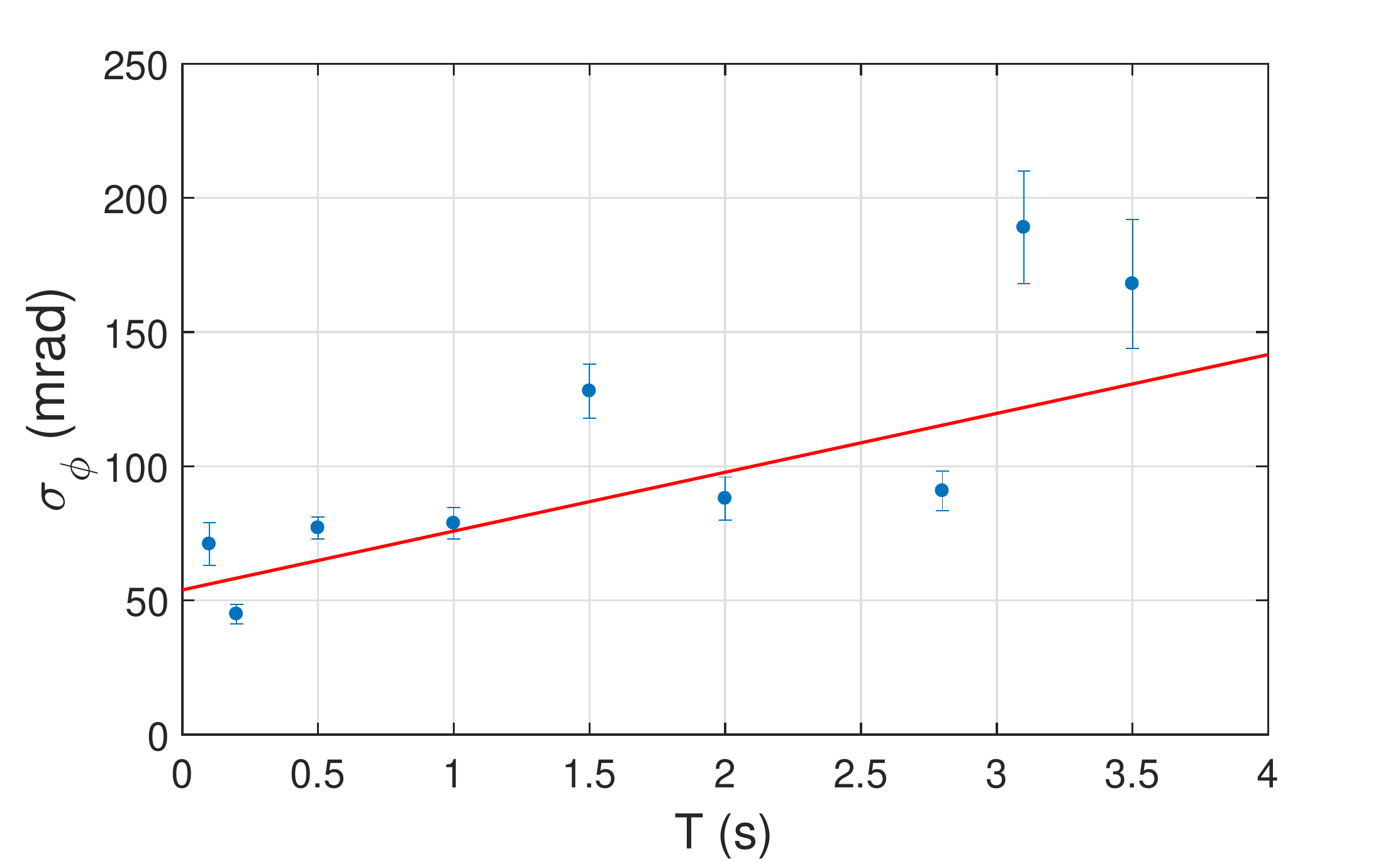}
  \caption{Interferometer phase noise $\sigma_\phi$ as a function of the interferometer time $T$. The solid red line is a linear fit to the data.}
  \label{PhaseNoiseT} 
\end{figure}

To determine the interferometer phase noise, we operate the interferometer at mid fringe and calculate the Allan standard deviation of the measured transition probability. To separate phase noise from detection noise, we then (quadratically) substract this latter contribution (estimated from \eref{sigma_det} at a given atom number).
The results, displayed in \fref{PhaseNoiseT}, show an increase of the measured interferometer phase noise as a function of $T$. To obtain an analytic expression and take this increase of the phase noise with $T$ into account in the estimation of the relative sensitivity (see next section), we perform a linear fit to the data with the law $\sigma_\phi=\sigma_{\phi_0} + k\, T$. We find $\sigma_{\phi_0}$ = (56 $\pm$ 7) mrad and $k$ = (21 $\pm$ 6) mrad/s. The dispersion of the data could be due to non stationarity of the measured noise, the graph collecting measurements realized over several days. We verify that using a quadratic fit to the data or even a simple time-independant average of the phase noise does not change significantly the optimal parameters of the interferometer derived in the following.

\subsection{Expected relative sensitivity}

Now that we have determined how the paramaters C, $\sigma_\phi$ and $\sigma_{det}$ depend on $T$ and $N_{at}$, we can evaluate the expected sensitivity for different trap geometries and number of atoms. We use equations \eref{sigma_nu}, \eref{sigmaP_mes} and \eref{sigma_det}, the fit from figure \ref{decayB} for the expression of $C(N_{at},T)$ and the fit from figure \ref{PhaseNoiseT} for the expression of $\sigma_{\phi}(T)$.

\begin{figure}[!ht]
  \centering
  \includegraphics[width=1\textwidth]{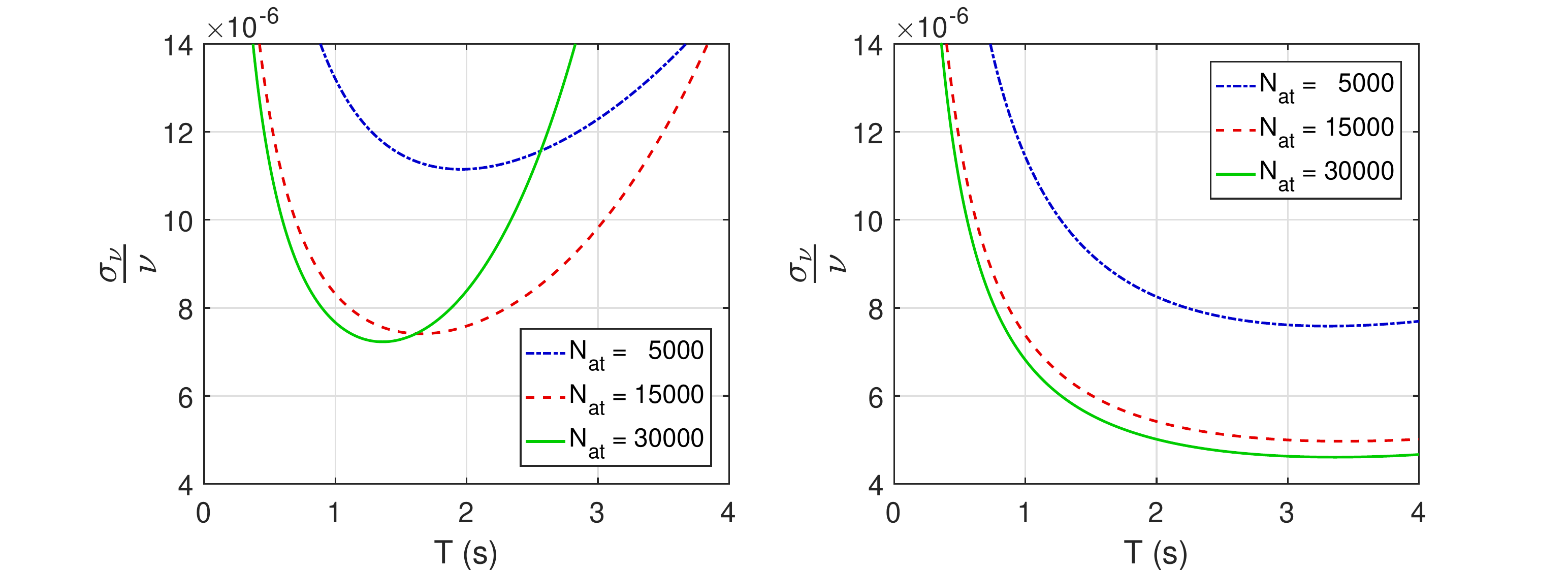}
  \caption{Expected relative sensitivity at 1\,s as a function of the interferometer time, for three different atom numbers, calculated with equations \eref{sigma_nu} and \eref{sigmaP_mes}, where the contrast, the detection noise and the phase noise are extracted from measurements (see text). Left: Transverse confinement laser waist = 150 $\upmu$m  --  Right: Transverse confinement laser waist = 300 $\upmu$m.}
  \label{sensibCalc} 
\end{figure}

\Fref{sensibCalc} displays calculated short term relative sensitivities $\frac{\sigma_\nu}{\nu}$ as a function of $T$ for three numbers of atoms and two IR transverse waist sizes (left: smaller waist = 150\,$\upmu$m, right: larger waist = 300\,$\upmu$m).
As expected, we find optimal interferometer times, which result from the compromize discussed above. This optimal interferometer time, and the corresponding optimal sensitivity, depends on the number of atoms.
Increasing the IR transverse size improves the optimal sensitivity, and reduces the variation of the optimal interrogation time with the number of atoms. 
Finally, optimizing on both parameters at a time, we find best sensitivities of the order of $\sim 5\times 10^{-6}$ at 1\,s for $T$ ranging from 3 to 3.5 seconds and for a number of atoms lying in between 10\,000 and 60\,000.

\subsection{Results}\label{results}

\begin{figure}[!ht]
  \centering
  \includegraphics[width=0.55\textwidth]{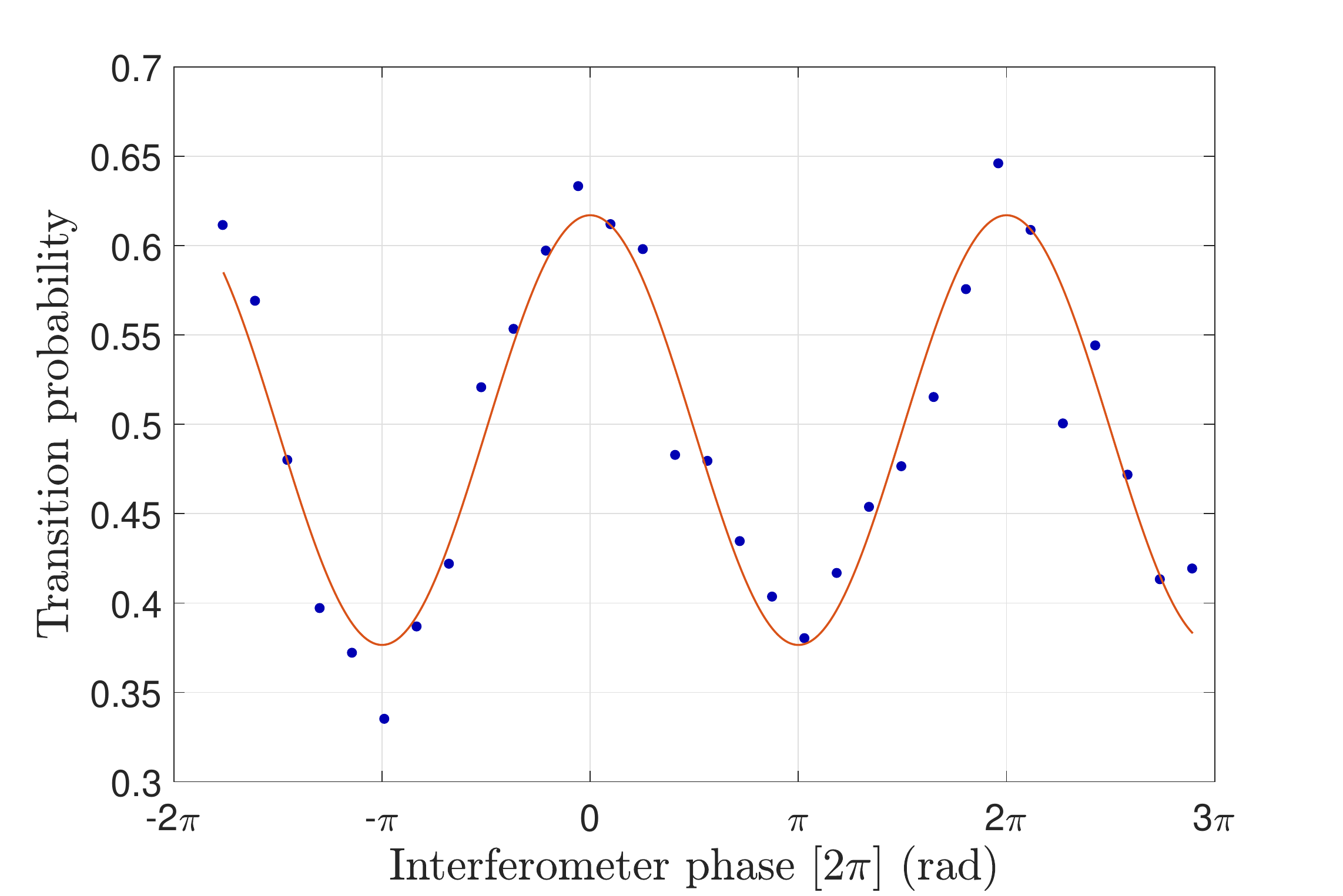}
  \caption{Typical Ramsey fringes for the optimized set of parameters $T$ = 2.7\,s and $N_{at}$ = 15000. The red line is a sinusoïdal fit to the data.}
  \label{fringes}
\end{figure}

\begin{figure}[!ht]
  \centering
  \includegraphics[width=0.9\textwidth]{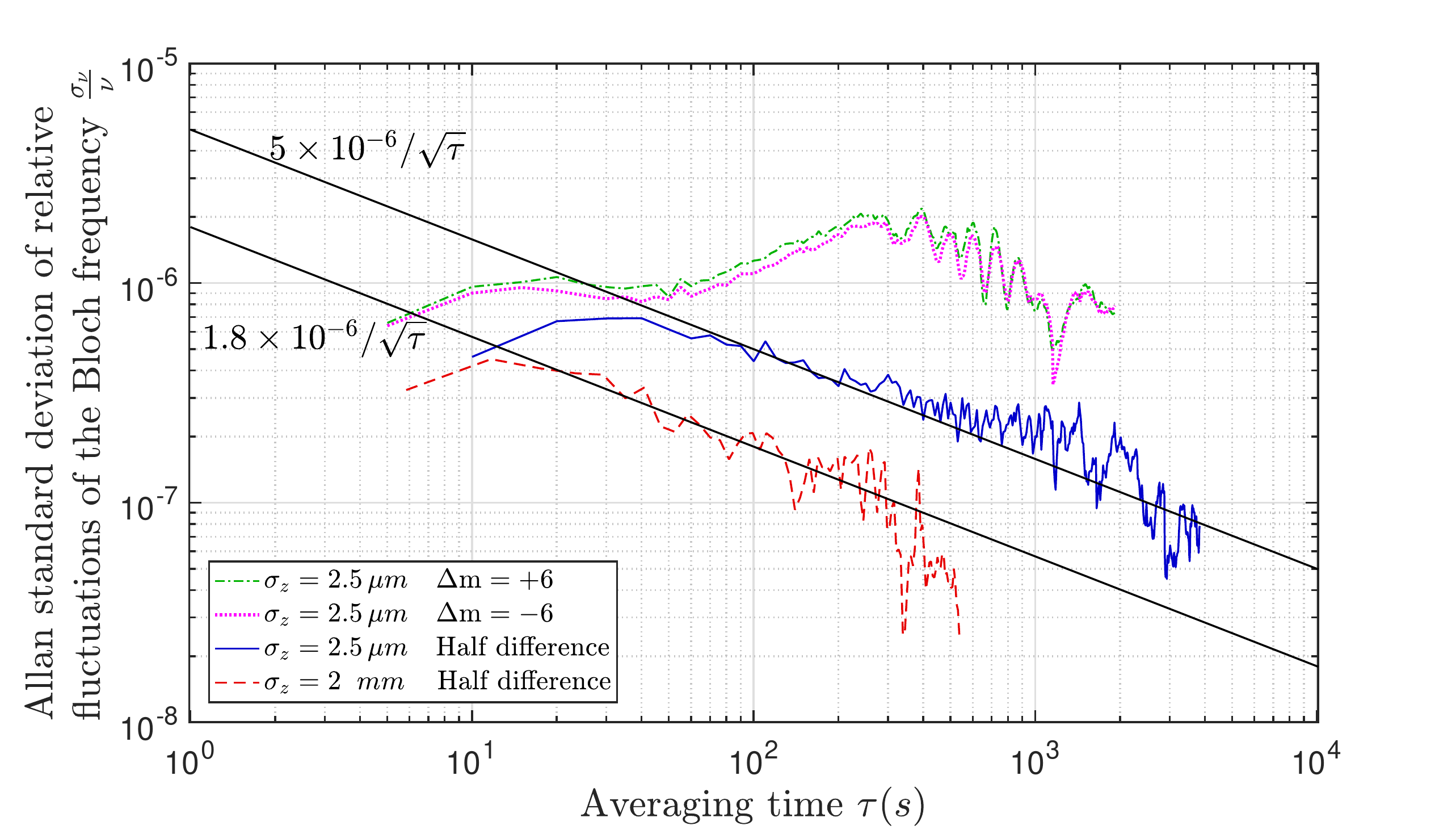}
  \caption{Allan standard deviation of the relative frequency fluctuations of the Bloch frequency measured with a spatial resolution of a few $\upmu$m with a cloud loaded from the optical dipole trap (DT) compared to the measurement with a cloud loaded from a molasses. The dotted pink (respectively dotted-dashed green) line corresponds to a measurement with a separation of $\Delta m$ = -6 (respectively +6) wells and a cloud loaded from the DT. The solid blue line is the differential measurement obtained from half the difference between these two measurements. The dashed red line is the measurement with a cloud loaded from a molasses (low resolution of 2 mm).}
  \label{AllVar}
\end{figure}

We then performed interferometer measurements for various $T$ and number of atoms, in the range 50 ms to 4 s and 1500 to 30000 atoms and found an optimal sensitivity of  $5\times10^{-6}$ for $T$ = 2.7\,s, $N_{at}$ = 15000 and $C$ = 0.3, close to the expected values.
A typical fringe pattern for such a set of parameters is shown in \fref{fringes}.
\\
\Fref{AllVar} displays the Allan standard deviation of the relative fluctuations of the Bloch frequency measurements.
The dotted pink line corresponds to a measurement with a separation of $\Delta m$ = -6 wells and the dotted-dashed green line with a separation of $\Delta m$ = +6 wells. We observe on both measurements a bump at about 5 minutes averaging time, which correspond to  half the period of the air conditioning system. The corresponding temperature cycles result in alignement and polarization fluctuations whose main effects are atom number and atomic density fluctuations on the one hand, and Raman coupling fluctuations on the other hand.
The solid blue line is the differential measurement corresponding to the half difference between interleaved $\Delta m$ = -6 and +6 measurements. This method rejects common mode frequency shifts and supresses the drifts discussed above. The relative stability of the differential measurement decreases down to $8\times10^{-8}$ after one hour of integration.
To compare with earlier results \cite{SolaroThesis}, the measurement with a cloud loaded from a molasses (much lower density and spatial resolution of about 2\,mm) is also presented (dashed red line). It reached a short term relative sensitivity of $1.8\times10^{-6}$ at 1\,s. This best sensitivity was achieved thanks to a shorter dead time (no evaporation stage), and to a negligible impact of interactions as already explained.

\subsection{Limitations}
\label{noise}

To better understand the limits in the sensitivity, we have performed an exhaustive analysis of the impact of all noise sources.
The different contributions are listed in \tref{noises}. They are expressed as a shot to shot frequency noise, for our cycle time of 5\,s, and as contributions to the relative short term sensitivity at 1\,s measurement time.
All the following results are given for the set of parameters that optimizes the sensitivity: $C = 0.3$, $T = 2.7$ s and $N_{at} = 15000$.

\begin{table}[!ht]
  \centering
\begin{tabular}{|l|c|c|}
  \hline
  Noise & $\sigma_{\nu}$ shot to shot (mHz) & $\sigma_{\nu_B}/\nu_B$ at 1 s \\
  \hline			
  Detection 		& 5.1		& $3.34 \times 10^{-6}$		\\
  Vibrations 		& 2.65		& $1.7 \times 10^{-6}$		\\
  Reference signal	& 1.32		& $0.87 \times 10^{-6}$		\\
  Trapping lasers	& 1.7		& $1.11 \times 10^{-6}$		\\  
  \hline
  sum 				& 6.14 		& $4 \times 10^{-6}$		\\
  \hhline{|=|=|=|}
  Interferometer	& 7.6		& $5 \times 10^{-6}$		\\
  \hline
\end{tabular}
\caption{Limitations to the sensitivity on the Bloch frequency measurement for an interferometer time $T = 2.7$ s, $N_{at} = 15000$ atoms and a contrast $C = 0.3$ \label{noises}}
\end{table}

The main contribution is the detection noise and has been described in section \ref{sec:sigmaP}.

\subsubsection*{Vibrations}
\leavevmode\par
Since we are probing the vertical potential, our sensor is sensitive to inertial noise in this direction.
The acceleration due to the vibration of the lattice retroreflecting mirror is measured with a seismometer placed at the top of the vacuum chamber, next to the mirror. The velocity signal is acquired during the interferometer and weighted by the transfer function of the interferometer. We calculate frequency fluctuations due to the vibrations of $\sigma_{\nu,vib}\sim$ 2.65 mHz.

\subsubsection*{Phase noise of the reference signal}
\leavevmode\par

A crystal oscillator is used as a reference for the frequency of the microwave source and for the frequency difference betwen the Raman lasers.
We use an ultra low phase noise oven controlled crystal oscillator (O-CDFF28ISN-R-10MHz/100MHz from NEL Frequency Controls, Inc.). To avoid long term drifts, the quartz is locked onto an ultra stable reference signal distributed in the laboratory, with a bandwidth of the order of 0.1\,Hz. The power spectral density of phase fluctuations of our quartz oscillator was measured, and the impact of its phase noise onto the symmetrized Ramsey-Raman interferometer phase noise was calculated, using the formalism of the sensitivity function \cite{Dick1987,Cheinet2008}. For our interferometer parameters  ($T_p$ = 5\,s and $T$ = 2.7\,s), we obtain a shot to shot frequency fluctuation due to our reference signal of $\sigma_{\nu,ocxo}$ = 1.32 mHz.

This contribution is not directly measurable since we cannot seperate it from other noise contributions such as the trapping laser fluctuations.

\subsubsection*{Trapping lasers}
\leavevmode\par
The trapping laser (lattice laser and transverse confinement laser, see \sref{sec:preparation}) intensity fluctuations induce differential light shift fluctuations which increase with the laser power.
To evaluate their impact, we realize a symmetrized Ramsey-MW interferometer with $T=3$\,s using only MW pulses (see figure \fref{RMWS}) in order to be insensitive to Raman coupling fluctuations. We measure a shot to shot frequency noise of 2.3\,mHz with 17000 atoms. We then substract the detection noise and the noise due to the reference signal calculated above, we then find a noise of 1.7\,mHz that we attribute to the trapping laser intensity fluctuations.

\subsubsection*{Raman laser}
\leavevmode\par
Raman lasers can bring additional contributions to the noise budget : laser phase noise (especially outside the bandwidth of the PLL) and differential light shift fluctuations. To evaluate their impact, we compared the sensitivities of symmetrized Ramsey interferometers with MW pulses and with Raman pulses, and found no difference. We thus found that the Raman lasers do not add a significant noise contribution.

\subsubsection*{Lattice depth}
\leavevmode\par
When driving the symmetrized Ramsey-Raman interferometer (see figure \fref{RRSMW}), the Raman coupling between the ground and excited state depends on the lattice depth \cite{Tackmann2011}. Thereby, lattice laser fluctuations (intensity, waist, pointing), as well as pointing instabilities of the transverse confinement laser, induce coupling fluctuations through the depth variations seen by the atoms. Even if the phase of the interferometer is in principle insensitive to coupling fluctuations, these modify the contrast and offset of the fringe pattern, and hence the transition probability.
We will thus in the following determine the amplitude of depth fluctuations, and quantify their impact on the measurement of the transition probability.

First, we drive a $\pi$/2 Raman pulse on the $\Delta m = 6$ transition at a lattice depth of $\sim$ 2.5\,E$_{rec}$, away from the optimal depth of 1.9\,E$_{rec}$ and where the coupling varies linearly with the depth \cite{Beaufils2011,SolaroThesis}. We first determine this slope by measuring the change of transition probability when deliberately varying the depth from 2.3 to 2.7 E$_{rec}$. Then we set the depth to 2.5 E$_{rec}$ and measure the fluctuations of the transition probability. This last measurement, combined with the slope, allows to evaluate the amplitude of shot to shot depth fluctuations of about 1\%. This results in relative coupling fluctuations as low as $\sigma_{\Omega}/\Omega \sim 5\times10^{-4}$ when the depth is adjusted for optimal coupling. The impact of such coupling fluctuations is calculated to be negligible (on the order of 10$^{-5}$\,Hz shot to shot fluctuations).

\subsubsection*{Conclusion}
\leavevmode\par

The quadratic sum of these different contributions is close to the fluctuations $\sigma_P$ measured with the optimized set of parameters. The slight difference could be explained by non-stationarity in the noise, by additional light shift induced by stray light from the dipole trap laser or eventually by an unidentified source of noise.

\section{Prospect for short range forces measurement}

Operating this sensor close to the lattice retro-reflecting mirror will allow to probe for short range forces. The atoms will be moved in the vicinity of this surface by the mean of a moving lattice.
The resolution could be improved further by selecting the atoms in a single Wannier-Stark state, by lifting the degeneracy between neighbouring transitions. This could be realized for instance by manipulating Zeeman states in a magnetic field gradient \cite{Schrader2004}, by applying an additional light shift or force using an appropriately shaped optical potential \cite{Wang2015}, or even using the Casimir-Polder interaction itself when very close to the surface of the mirror. 

\begin{figure}[!ht]
  \centering
  \includegraphics[width=1\textwidth]{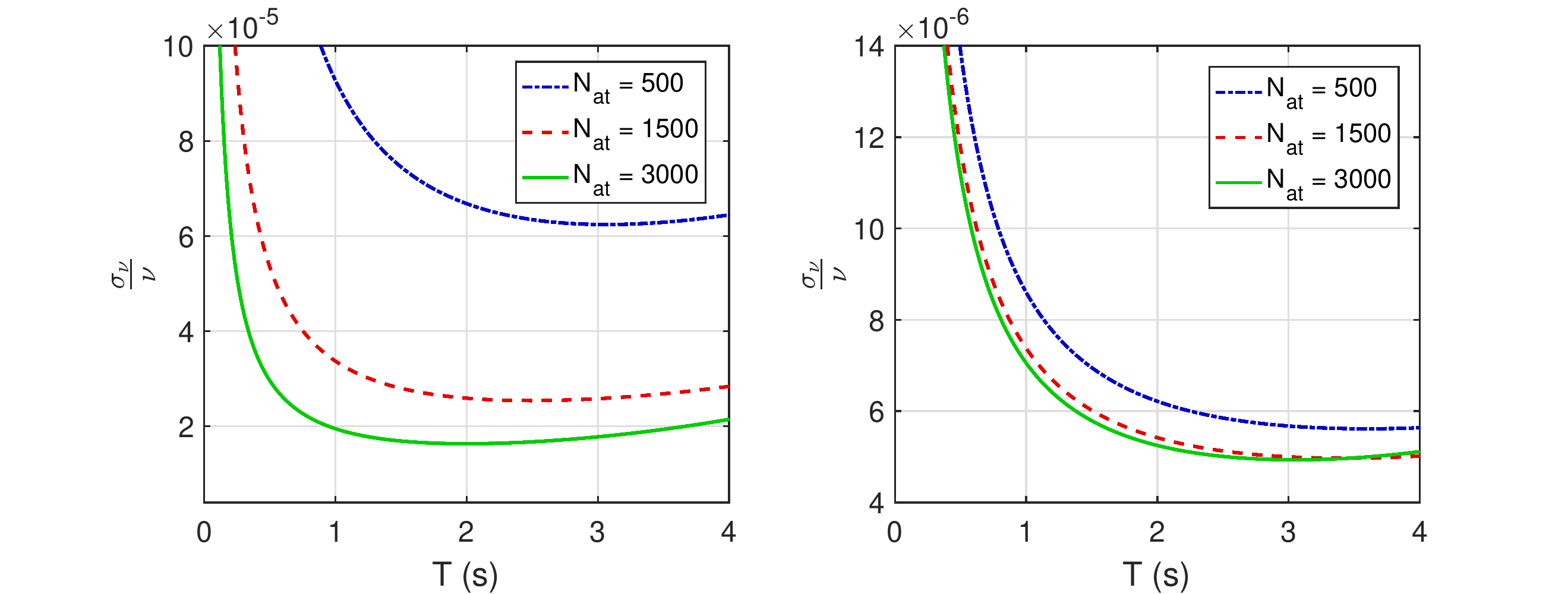}
  \caption{Left: Expected relative sensitivity after selection, assuming 10\% of the atoms are selected.  --  Right: Same as left but with a detection noise reduced down to the quantum projection noise level.}
  \label{sensibCalc_selection} 
\end{figure}

The number of atoms will then be reduced, thus increasing detection noise and degrading the sensitivity, as presented on the left plot of \fref{sensibCalc_selection}, assuming a selection of 10\% of the initial sample. Nevertheless, improving the detection noise down to the quantum projection noise level would allow to maintain the sensitivity, as shown on the right plot of \fref{sensibCalc_selection}.

\section{Spatial resolution}\label{resolution}

We define the spatial resolution of our sensor as  the time averaged standard deviation of the atomic position distribution along the vertical axis $\sigma_z$. 
As the resolution of our imaging system is about 4.5\,$\upmu$m, we cannot measure this distribution purely optically. We will thus infer the spatial resolution from the determination of the mean atomic density and radial size of the sample. The mean density will be deduced from the measurement of the frequency shift induced by cold collisions in a standard Ramsey-MW interferometer performed in the trap. It is given by \cite{Sortais2000}:
\begin{equation}
\Delta_{\nu} = \frac{2\hbar}{m_{Rb}}(a_{22}-a_{11})\bar{n}
\label{CollShift}
\end{equation}
where $a_{11}$ and $a_{22}$ are the relevant scattering lengths.

\subsection{Atomic distribution in a periodic potential}
\begin{figure}[!ht]
  \centering
  \includegraphics[width=1\textwidth]{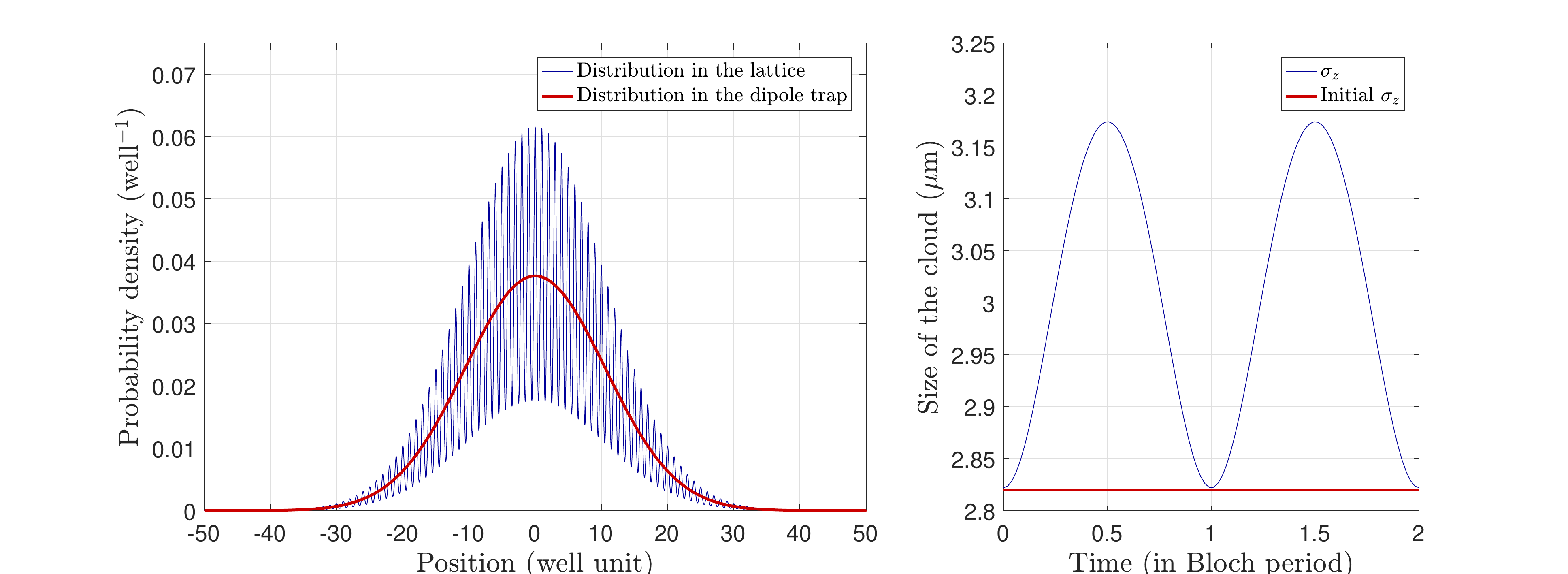}
  \caption{Probability density in the vertical direction in the harmonic dipole trap and in the lattice at a depth of 1.9\,E$_{rec}$ (left) with the corresponding size over two Bloch periods (right).}
  \label{size}
\end{figure}

A knowledge of the atomic distribution is necessary to link the measured mean density to $\sigma_z$. 
To determine this distribution, we first model the state of the atomic sample after the evaporation as a statistical mixture of minimal wavepackets, of temperature 300 nK, distributed in a Gaussian distribution, of rms radius $\sigma_{z,DT}$. 
The initial quantum state in the lattice is then obtained by projecting these wavepakets into the subbasis of the WS eigenstates of the fundamental band of the lattice. The statistical mixture of projected wavepackets leads to the atomic distribution displayed as a blue line in the left part of \fref{size}. 
There, the initial Gaussian distribution with $\sigma_{z,DT}=2.8\,\upmu$m is displayed as a red line. The lattice depth is 1.9\,E$_{rec}$. 
The position distribution in the lattice is not stationary, but evolves periodically at the Bloch frequency. The corresponding evolution of $\sigma_z$ is displayed at the right of \fref{size}.
Having determined the distribution, we can now calculate the mean linear vertical density and the rms size $\sigma_z$.
We finally find the following relationship between these two quantities (at the lattice depth of 1.9\,E$_{rec}$):
\begin{equation}
\kappa = \bar{n}_z \sigma_z = 0.34
\label{kappa}
\end{equation}
We find that this product does not depend on the initial cloud size $\sigma_{z,DT}$ as long as this size is larger than two lattice sites. More, its value varies by less than 3\% during a Bloch oscillation, since the size increases while the density decreases. The mean density we measure is in fact averaged over a Bloch period, thus corresponding to a time averaged size. For the example given in \fref{size}, the size $\sigma_z$ oscillates between 2.8 and 3.2\,$\upmu$m (right).

The trap being harmonic in the tranverse directions, the vertical size of the cloud can now be deduced from the average density $\bar{n}_{3D}$:
\begin{equation}
\sigma_z = \kappa\,\frac{N_{at}}{4\pi\sigma_r^2\bar{n}_{3D}}
\label{sigmaz}
\end{equation}

\subsection{Results}

With a waist of 300\,$\upmu$m and a power of 1.36\,W for the transverse confinement laser, we measure with our imaging system a transverse size of $\sigma_r$ = (74 $\pm$ 5) $\upmu$m.

The collisional frequency shift is measured to be (25 $\pm$ 5) mHz for 30000 atoms, from which we deduce, with equation \ref{CollShift}, a density of $\bar{n}$ = (5 $\pm$ 1)$\times 10^{10}$ at/cm$^3$.\\
The corresponding vertical size, derived from equation \ref{sigmaz}, is then $\sigma_z$ = 3.0 $\pm$ 0.7 $\upmu$m, at our lattice depth of 1.9\,E$_{rec}$.
This corresponds to an initial size $\sigma_{z,DT}$ of 2.8\,$\upmu$m as chosen in \fref{size}.

Here, the confinement in a shallow lattice increases the parameter $\kappa$ (of equation \ref{kappa}) by a factor 1.2 only, with respect to the harmonic case where $\kappa_{harmo} = \frac{1}{2\sqrt{\pi}}$. For deeper lattices, the effective volume is significantly smaller and the density increases. At 10\,E$_{rec}$, we calculate $\kappa$ to increase by a factor of about 2.

The size $\sigma_z = 3\,\upmu$m determined here is the one for which we obtained the best sensitivity. Evaporating further down resulted in a drastic loss of atoms and a degradation of sensitivity. With a different dipole trap geometry (with waists of 25 and 200 $\upmu$m), we obtain smaller sizes of about 1\,$\upmu$m, but with higher densities in the $\sim 10^{12}$ at/cm$^3$ range, resulting in a degraded optimal relative sensitivity.

\subsection{Impact of the cloud center oscillations}

In principle, the spatial resolution determined above is degraded by the periodic motion of the center of the cloud (Bloch oscillations in position). At a depth of 1.9\,E$_{rec}$, the amplitude of this oscillation is of about 0.6\,$\upmu$m.
When averaged over one cycle, this results, in our case, in a negligible increase (less than 1\%) of the width of the position distribution. This change would be more pronouced for smaller initial clouds. For example, in the limit of a single wavepacket at 300\,nK, we calculate an increase of 15\%.

\subsection{Delocalization and Casimir-Polder measurement}
We now discuss the impact of the delocalization of the wavefunction on the measurement of the Casimir-Polder (CP) interaction. The shifts of the energy levels of the WS ladder differ from the simple expression of the CP potential at each energy site, and need to be precisely calculated. Such calculations were performed in \cite{Maury2016,Messina2011}, where the influence of the presence of the surface and of the CP interaction on the eigenstates and eigenenergies of the problem have been numerically evaluated. 
In principle, comparing the results of \cite{Maury2016} with the measurements to come will require a precise knowledge of the distribution of the WS states initially populated, except if only a single WS state is populated or selectively interrogated. The wavefunction will nevertheless remain spatially delocalized at shallow lattice depths (at 1.9\,E$_{rec}$, the rms size of the WS states is 3 lattice sites).
The impact of this delocalization onto the phase shift of the interferometer could be reduced, for instance by increasing the lattice depth during the free evolution period.


\section{Conclusion}

We have demonstrated a quantum force sensor which combines both a very high spatial resolution of 3\,$\upmu$m and a very high sensitivity of $5\times10^{-6}$, and where interaction-induced dephasing is efficiently mitigated by diluting the atomic sample in the transverse direction.

\begin{table}[!ht]
  \centering
\begin{tabular}{|c|c|c|c|c|}
  \hline
   						& Resolution	& Relative sensitivity  		& 		$\eta$				& Relative sensitivity\\
   						&				& on g							&							& on Casimir-Polder\\
  \hline
  This work				& 3 $\upmu$m	& $5 \times 10^{-6}$ at 1\,s	& $1.5\times10^{-11}$\,m	& 1\% after 5\,s\\
   						&				&								&							& averaging time\\
  \hline
  \cite{Tarallo2014}	& 37 $\upmu$m	& $1.5 \times 10^{-6}$ at 1\,s	& $5.6\times10^{-11}$\,m	& N/A			\\
  \hline
  \cite{Sorrentino2009}	& 12 $\upmu$m	& $5 \times 10^{-6}$ at 1\,s	& $6\times10^{-11}$\,m		& N/A			\\
  \hline
  \cite{Harber2005}		& 2.4 $\upmu$m	& N/A							& N/A						& 10\% after 10\,min\\
  						&				&								&							& averaging time\\
  \hline
\end{tabular}
\caption{Comparison between a selection of trapped force sensors \label{comparaisons}}
\end{table}

To establish a figure of merit for our atomic interferometer, we define the variable $\eta = \frac{\sigma_{\nu}}{\nu}\sigma_z = \frac{\sigma_F}{F}\sigma_z$, where $\frac{\sigma_{\nu}}{\nu}$ is the relative sensitivity at 1s (see equation \eref{sigma_nu}), $F$ is the measured force corresponding to the frequency $\nu$ and $\sigma_z$ is the width of the Gaussian vertical atomic distribution, assimilated to the spatial resolution. With the size of $\sigma_z = 3\,\upmu$m (see \sref{resolution}) and our best sensitivity at 1 s of $5\times10^{-6}$, we reach a value of $\eta = 1.5\times10^{-11}$\,m. The force sensor described in \cite{Sorrentino2009} is based on the measurement of Bloch oscillations of $^{88}$Sr atoms trapped in a vertical lattice. A resolution of $\sigma_z = 12\ \upmu$m is achieved, combined with a relative sensitivity of $5\times10^{-6}$, thus corresponding to $\eta = 6\times10^{-11}$. In \cite{Tarallo2014}, a state-of-the-art relative sensitivity of $1.5\times10^{-6}$ is obtained with the same experiment but the measurement is based on the delocalization of the wave-packet and thus can't achieve the same resolution, the cloud width reaches 37\,$\upmu$m corresponding to $\eta = 5.6\times10^{-11}$\,m.

Well resolved measurements of the Casimir-Polder (CP) potential using ultracold atoms were performed in \cite{Harber2005}. In this experiment, the CP potential is derived from the shift in the oscillation frequency of the center of mass of a BEC trapped in a magnetic field in the vicinity of a surface. A Thomas-Fermi radius of $\sigma_z = 2.4\,\upmu$m is achieved and the sensitivity on the frequency shift allows to reach a relative sensitivity on the CP potential at 6$\upmu$m from the surface of about 10\%. This uncertainty is obtained after more than ten minutes measurement time.
In comparison, the sensitivity we have demontrated would allow for the determination of the CP potential at the same distance of 6\,$\upmu$m with an uncertainty of the order of 1\% in a single shot measurement.

This combination of a very high sensitivy on a force measurement and of a very high spatial resolution makes this sensor a perfect device for short range forces measurements such as the CP force as mentionned above or for the search of a deviation to the gravitational potential at short range \cite{Wolf2007,Adelberger2003}.

\section{Acknowledgement}

We acknowledge financial support by the IDEX PSL (ANR-10-IDEX-0001-02 PSL) and ANR (ANR-13-BS04-0003-01). A. Bonnin thanks the Labex First-TF for financial support.

\pagebreak
\bibliographystyle{unsrt}
\bibliography{Bibliography}

\end{document}